\documentclass[graybox]{svmult}

\usepackage{mathptmx}       % selects Times Roman as basic font
\usepackage{helvet}         % selects Helvetica as sans-serif font
\usepackage{courier}        % selects Courier as typewriter font
\usepackage{type1cm}        % activate if the above 3 fonts are
\usepackage{makeidx}         % allows index generation
\usepackage{graphicx}        % standard LaTeX graphics tool
\usepackage{multicol}        % used for the two-column index
\usepackage[bottom]{footmisc}% places footnotes at page bottom
\usepackage{cite}
\makeindex             % used for the subject index
\begin{document}

\title*{Magnetic phase transitions in the Ising model}
\author{P.D.Andriushchenko and K.V.Nefedev}
\institute{P.D.Andriushchenko \at Far Eastern Federal University, School of Natural Sciences, 8 Sukhanova St., Vladivostok 690950, Russia.
 \email{pitandmind@gmail.com}
\and K.V.Nefedev \at Far Eastern Federal University, School of Natural Sciences, 8 Sukhanova St., Vladivostok 690950, Russia. \email{knefedev@phys.dvgu.ru}}
\maketitle

\abstract*{In this paper we consider an approach, which allows researching a processes of order-disorder transition in various systems (with any distribution of the exchange integrals signs) in the frame of Ising model. A new order parameters, which can give a description of a phase transitions, are found. The common definition of these order parameters is the mean value of percolation cluster size. Percolation cluster includes spins with given energy. The transition from absolute disorder to correlated phase could be studied with using of percolation theory methods.}

\abstract{ In this paper we consider an approach, which allows researching a processes of order-disorder transition in various systems (with any distribution of the exchange integrals signs) in the frame of Ising model. A new order parameters, which can give a description of a phase transitions, are found. The common definition of these order parameters is the mean value of percolation cluster size. Percolation cluster includes spins with given energy. The transition from absolute disorder to correlated phase could be studied with using of percolation theory methods.}

\section{Introduction}
\label{sec:1}
Ordered systems, such as ferromagnets are studied well now. Physics of systems with a complex type of exchange interaction is not so simple and evident. For instance, the theoretical research of the paramagnetic-spin glass transition, which started several decades ago, is still in  process   ~\cite{Vasin:2006,Ginzburg:1989}. The theory of magnetic states and transitions from paramagnetism to
antiferromagnetism of various types (A, B, C, CE, G, and others), for instance in manganites, is still in development ~\cite{Nagaev:1996,Nagaev:2001,Gor'kov:2004,Shuai:2008}. It is well known that for such systems the average magnetization cannot be used as an order parameter.  At low temperature the correlations between spins grow. This fact is proved by known temperature dependence of specific heat and magnetic susceptibility behavior and difference in the temperature behavior of the magnetization, which is measured in ZFC and FC modes.

In this paper, we present the result of research of magnetic phase transitions in the Ising model on a simple square lattice using the numerical simulation methods. We worked with the following three models: with ferromagnetic interactions, antiferromagnetic interactions and random distribution of exchange integrals (spin-glass models).

\section{Parallel algorithm for finding the equilibrium configuration}
\label{sec:2}

\begin{figure}[b]
\centering{\includegraphics[width=80mm]{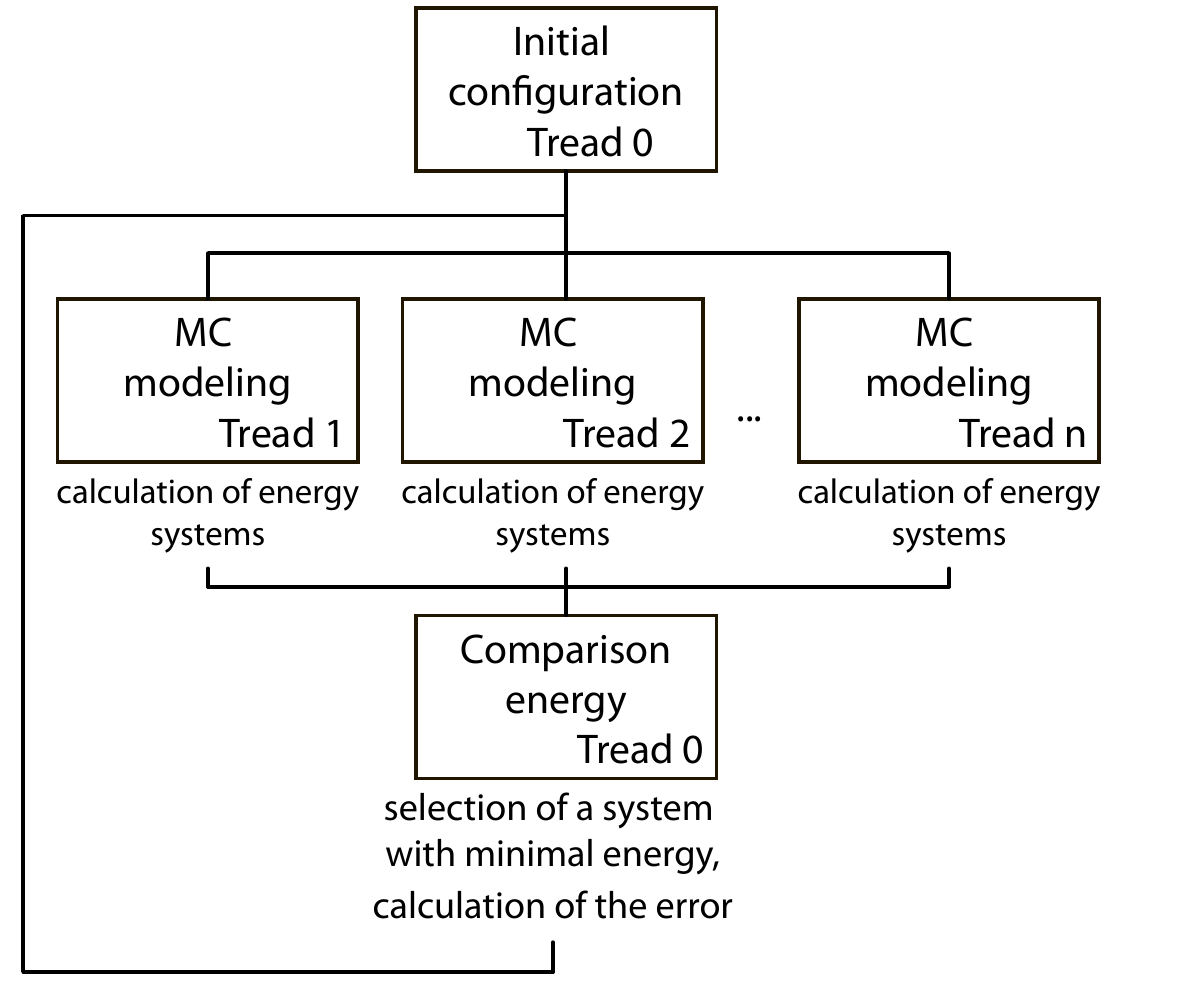}}
\caption{Scheme of the parallel realization of Monte Carlo algorithm}
\label{fig:1}       % Give a unique label
\end{figure}

  The parallel search scheme for the equilibrium configuration is shown in Fig.~\ref{fig:1}. The values of spins, their energies, as well as links of a square lattice of the magnet were recorded in one-dimensional dynamic arrays for more flexible allocation of memory. At the start of simulation the temperature is set to be corresponding to paramagnetic state (in reduced units $T = 4.5$, which is higher than the Curie temperature for a square lattice, obtained by Onsager $T_c = 2.28$). The initial configuration, that corresponds to the random distribution, of spin directions, is generated. Configuration data, the temperature, the number of MC steps and other technical information are sent out to all the computing processes by means of MPI technology.
 
Each process performs a Monte Carlo spin flip (approx. $10 ^ 9$  spin flips for $1000\times1000$  spin system) until the system comes into equilibrium, which is defined by the energy of the system. The system came into equilibrium at the temperature $T$ in case if the system's energy became less after certain quantity of MC steps (for modeling the transition with decreasing temperature) and will not change significantly during further modeling. The energy of configurations at the given temperature are passed into the root thread, which one also compares the obtained values. Root process selects the energy corresponding to the extreme value (for modeling the transition with decreasing temperature -- minimum value). All variables are recorded to the output files, and the temperature lowered at the defined
value ($\Delta T = 0.1$). The thread, which number is defined by the root process and with the lowest system energy, distributes the data about the selected equilibrium distribution of spins and the new temperature value to the other threads.
This cycle is repeated until the temperature achieve zero. The average magnetization is calculated by the difference between the spins directed upward and downward. The total energy is calculated by the summarizing all energies of the interacting spin pairs.

To count the number of nodes with minimal energy in the maximum cluster we can use breadth-first search (BFS) algorithm with the creation of queue (which is the graph traversal task).

\section{Ordering and bond-percolation on the simple square lattice }
\label{sec:3}

The probability of any possible configuration is given by the Gibbs distribution ~\cite{Landau:1980}. If we know the partition function for a system of interacting spins, it allows us to calculate all possible average physical values, which fully describe the state of the system under the given external conditions. Currently, in research of the phenomenon of percolation numerical methods is mainly used (Monte Carlo). They are widely used in statistical physics ~\cite{Newman:1999}. 

We only point out that in the Ising model with Hamiltonian

\begin{equation}
\label{01}
H=-\frac{1}{2}\sum \limits_{ij}{J_{ij}S_iS_j}
\end{equation}
which takes into account the ferromagnetic interaction, exchange integral $J_{ij} = 1$ between each spin $S_i$ and its nearest neighbor $S_j$. For antiferromagnetic interactions $J_{ij}= -1$.

\begin{figure}[t]
\centering{\includegraphics[width=90mm]{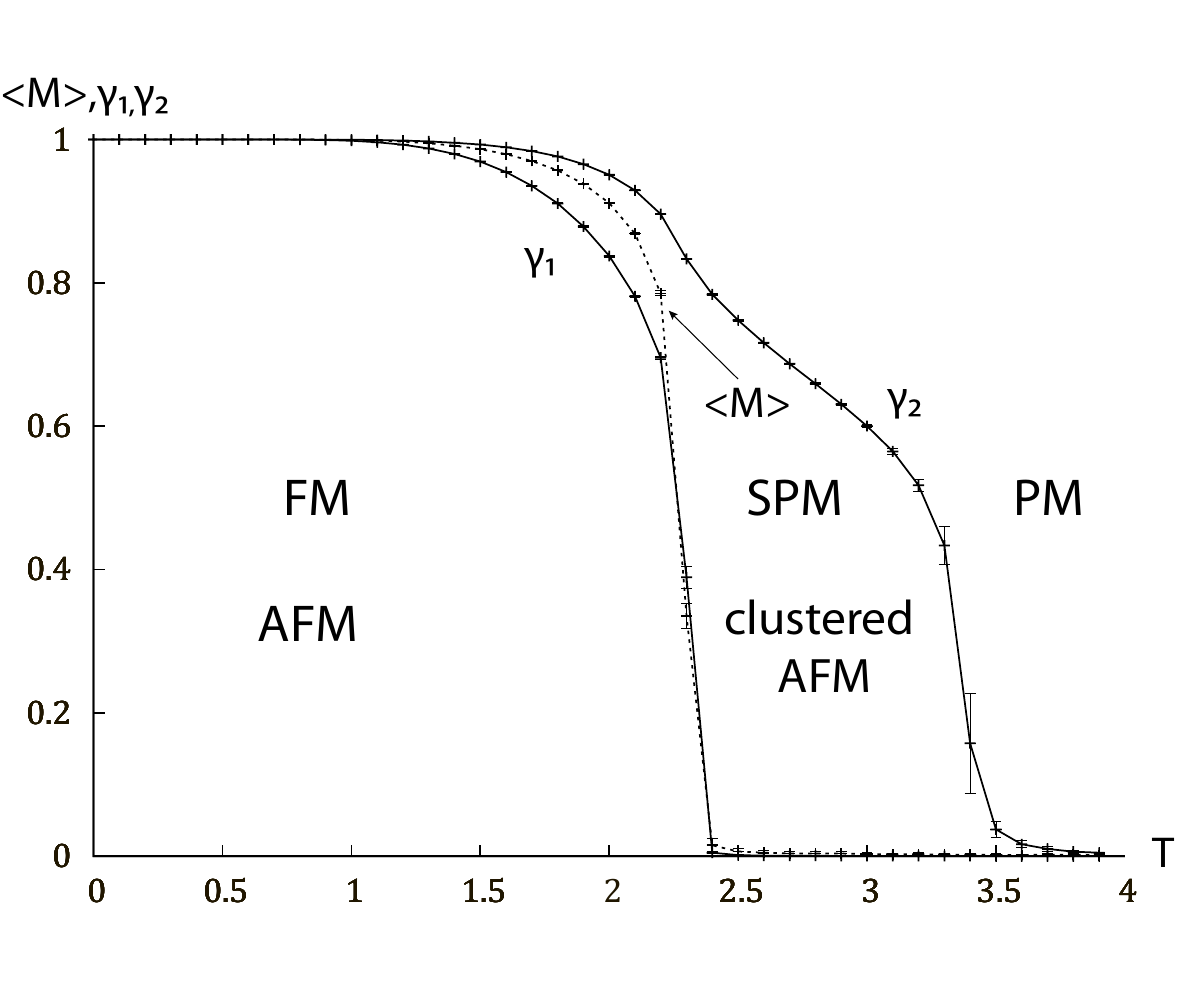}}
\caption{Temperature dependence of the relative size of the percolation cluster for Ising  $\gamma_1(T)$ with $J=+1$ and $J=-1$. The behavior of the relative number of spins $\gamma_2(T)$ in the maximal cluster (energy $E=-4$ and $E=-2$) and magnetization $<M(T)>$ (For antiferromagnetic $<M(T)>=0$ for any temperature). All values are presented in reduced units. FM - ferromagnetism, AFM - antiferromagnetism, SPM - superparamagnetism (clustered FM), PM - paramagnetism.}
\label{fig:2}
\end{figure}

Monte Carlo simulation with Metropolis algorithm allows to calculate the average relative magnetization $<M>$ of the ferromagnetic system $1000\times1000$ Ising spins (the number of Monte Carlo steps is $2 * 10^{10}$) on a simple square lattice with $z = 4$ nearest neighbors. Temperature behavior $<M>$ shown in Fig.~\ref{fig:2}.

The same method was used to calculate the average size of the percolation cluster $\gamma_1(T)$, which is defined as the ratio of the number of spins in the ground state to the total number of spins. There is a coincidence between the critical temperature of  the magnetization $<M>$ and the order parameter $\gamma_1(T)$. The law of variation and the critical exponents for the temperature dependence of the assumed physical value $\gamma_1(T)$ (average in time and configuration) coincide with observed characteristics for the average magnetization of ferromagnetic systems  ~\cite{Onsager:1944}. The difference in the temperature behavior within reviewed order parameters at $T \leq T_c$, (higher growing rate of $<M>$  compared to $\gamma_1$), caused by the fact that the percolation cluster does not contain all the spins in the ground state, that means the difference in growing rate caused by low density of the percolation cluster and extention of new phase clusters in its pores. Ordering process is usually characterized by the formation of a set of new phase nuclei. At $T = T_c$  only a part of small cluster are united in one most size the percolation cluster. Thus, in this model, the order parameter $<M>$ describes the balance between the number of particles "up" $N\uparrow$ and the number of particles "down" $N\downarrow$, and the proposed new order parameter $\gamma_1(T)$ describes the process of growth of the percolation cluster.

The results of this research show that $\gamma_1(T)$ in antiferromagnetic model (${J_{ij} = -1}$) have the same jump in the phase transition, as in the ferromagnetic model (${J_{ij} = +1}$), as shown in Fig.~\ref{fig:2}. While the magnetization $<M(T)>$ in antiferromagnetic systems is equal to zero at any temperature, and therefore cannot serve as an order parameter. This fact is due to the universality of order parameter  $\gamma_1(T)$.

Besides, the  function $\gamma_2(T)$ represents the relative size of the maximal cluster, which one unites a spins with the negative interaction energy. Fig.~\ref{fig:2} shows, that the function $\gamma_2(T)$ allows us to determine the transition temperature of the system from absolutely randomized paramagnetic (PM) phase to correlated superparamagnetic (SPM).

\section{Simulation of the transition from paramagnetism to spin glass}
\label{sec:4}
In 1975 S. Edwards and P. Anderson considered the lattice model of exchange-coupled magnetic moments interacting so that the exchange integral is a random function ~\cite{Edwards:1975}. In this spin glass the one half of pairs interacts ferromagnetically, and second part interacts antiferromagnetically. The types of interactions distributed randomly. The current research is based on the S. Edwardson - P. Anderson model specified to  frustration in every spin

\begin{equation}
\label{02}
\sum \limits_{i=1}^{z=4}{J_i=0}
\end{equation}
in case of simple square lattice (the summation is over $z=4$ neighbors).

\begin{figure}[t]
\centering{\includegraphics[width=90mm]{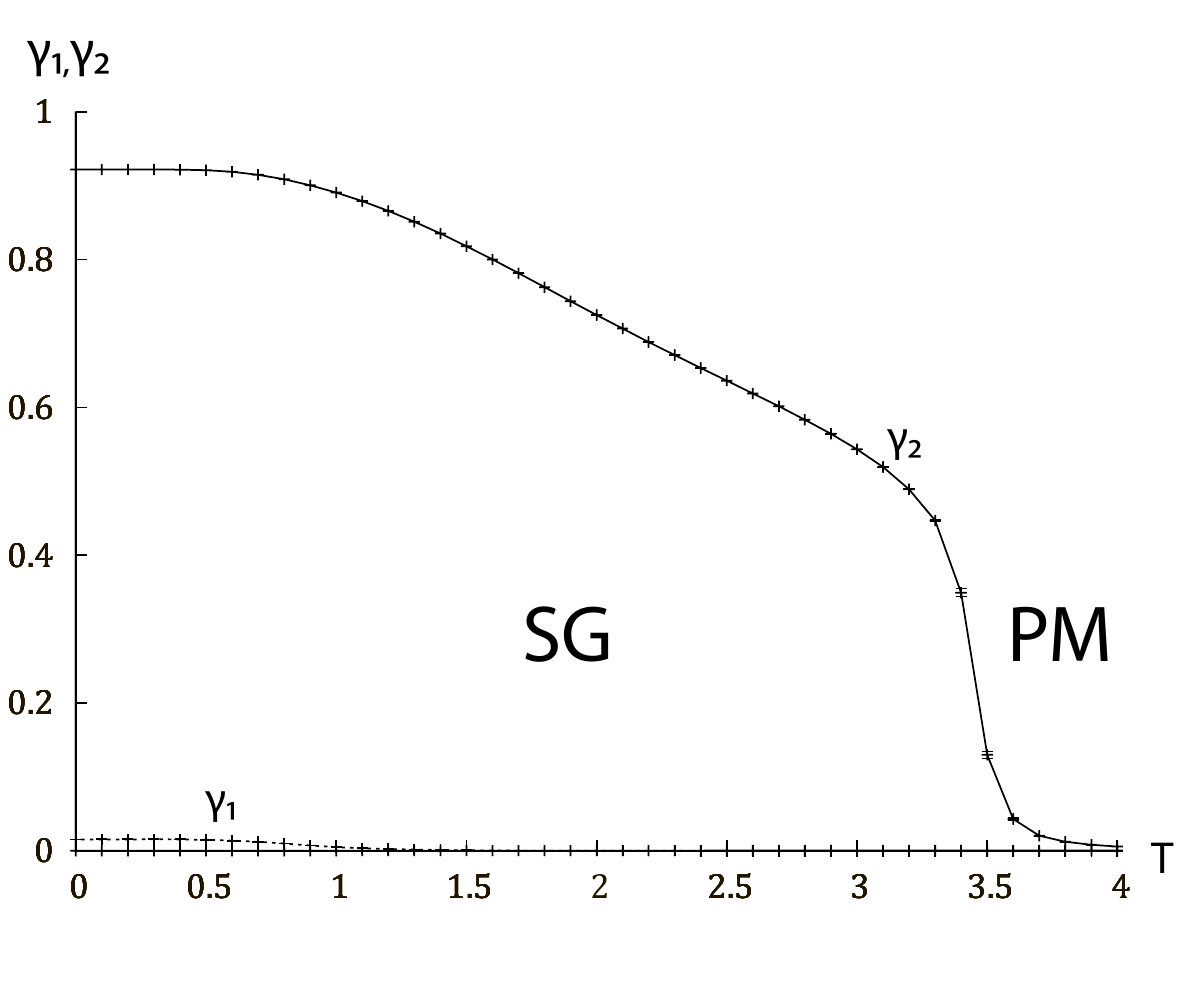}}
\caption{Temperature dependence of the relative size of the percolation cluster of Ising spins $\gamma_1(T)$  for spin glass model. Behavior of the relative number of spins $\gamma_2(T)$ in the maximal cluster (with energy $E= -4$ and $E= -2$). All values are presented in reduced units.}
\label{fig:3}       % Give a unique label
\end{figure}

Monte-Carlo simulation of transition processes in described above model of spin glass lead to the existence of specific critical temperature $T_f$, Fig.~\ref{fig:3}. In almost completely (99\%) frustrated system of $1000\times1000$ of Ising spin glass, there is the
function $\gamma_2(T)$ which one has abrupt changes of value in transition region. Function $\gamma_2(T)$ is not equal to $1$ even at $T=0$, because there is large number of frustrated spins in excited state. This typical phenomenon for spin glass state leads to nonzero magnetic capacity heat at zero value of the absolute temperature.

We suppose that in the limit of infinite number of particles this jump should be even more evident.

\section{Conclusion }
\label{sec:5}

The relative power of a percolation cluster could be used as a universal order parameter for systems with direct exchange interaction between spins. This parameter describes the short-range or of the long-range order, depending the ways of cluster integration of spins over values of exchange interaction energy. In the numerical experiments the possibility of phase separation of paramagnetic and superparamagnetic regimes, paramagnetic and spin glass states is showed. The results of given research can be summarized in the following conclusion:
\begin{enumerate}
\item The law of temperature behavior of the new order parameter -- function $\gamma_1(T)$ coincides with the law of temperature behavior of the average magnetization for a ferromagnet $<M>$. There is the coincidence critical temperature of magnetization formation and critical temperature of percolation threshold. 
\item The function $\gamma_1(T)$  in the antiferromagnet undergoes a jump at the point of the phase transition, and it's behavior is the same as $\gamma_1(T)$ in a ferromagnetic. It stands as the universality of the order parameter.
\item The existence of function which has the jump at the spin glass-paramagnetic transition region could give the solution of the phase transition problem in PM-SG (PM-SPM). The $\gamma_1(T)=0$ at spin glass tells about the absence of phase transition in this $2D$ lattice of Ising spins.
\item The proposed approach allows to unite the concepts of a phase transitions in ordered and disordered systems, including systems with competing interactions with the developed ideas in percolation theory.
\item The order parameter is universal for any magnetic system. It could be measured experimentally using spectroscopy methods.

\end{enumerate}

This approach can be extended to the case of a complex alternating-sign exchange of long-range interaction.

The interesting questions are the research of 3D lattice spin glass state for existence of phase transition and also the simulation ZFC and FC regimes.

This work was supported by grant No. 14.740.11.0289, No. 07.514.11.4013 and No. 02.740.11.0549 from Ministry of Education and Science of the Russian Federation.

\bibliographystyle{IEEEtran} 
\bibliography{citation1}  

\begin{thebibliography}{10}
\providecommand{\url}[1]{#1}
\csname url@samestyle\endcsname
\providecommand{\newblock}{\relax}
\providecommand{\bibinfo}[2]{#2}
\providecommand{\BIBentrySTDinterwordspacing}{\spaceskip=0pt\relax}
\providecommand{\BIBentryALTinterwordstretchfactor}{4}
\providecommand{\BIBentryALTinterwordspacing}{\spaceskip=\fontdimen2\font plus
\BIBentryALTinterwordstretchfactor\fontdimen3\font minus
  \fontdimen4\font\relax}
\providecommand{\BIBforeignlanguage}[2]{{%
\expandafter\ifx\csname l@#1\endcsname\relax
\typeout{** WARNING: IEEEtran.bst: No hyphenation pattern has been}%
\typeout{** loaded for the language `#1'. Using the pattern for}%
\typeout{** the default language instead.}%
\else
\language=\csname l@#1\endcsname
\fi
#2}}
\providecommand{\BIBdecl}{\relax}
\BIBdecl

\bibitem{Vasin:2006}
M.~Vasin, ``Description of the paramagnet-spin glass transition in the
  edwards-anderson model using critical-dynamics methods,'' \emph{Theoretical
  and Mathematical Physics}, vol. 147, pp. 721--728, 2006.

\bibitem{Ginzburg:1989}
S.L.Ginzburg, \emph{\BIBforeignlanguage{Russia}{Irreversible Phenomena in Spin
  Glasses}}.\hskip 1em plus 0.5em minus 0.4em\relax Nauka; Moscow, 1989, in
  Russian.

\bibitem{Nagaev:1996}
E.~L. Nagaev, ``Lanthanum manganites and other giant-magnetoresistance magnetic
  conductors,'' \emph{Physics-Uspekhi}, vol.~39, no.~8, pp. 781--805, 1996.

\bibitem{Nagaev:2001}
E.~Nagaev, ``Colossal-magnetoresistance materials: manganites and conventional
  ferromagnetic semiconductors,'' \emph{Physics Reports}, vol. 346, no.~6, pp.
  387 -- 531, 2001.

\bibitem{Gor'kov:2004}
L.~P. Gor'kov and V.~Z. Kresin, ``Mixed-valence manganites: fundamentals and
  main properties,'' \emph{Physics Reports}, vol. 400, no.~3, pp. 149 -- 208,
  2004.

\bibitem{Shuai:2008}
S.~Dong, R.~Yu, S.~Yunoki, J.-M. Liu, and E.~Dagotto, ``Ferromagnetic tendency
  at the surface of ce-type charge-ordered manganites,'' \emph{Phys. Rev. B},
  vol.~78, p. 064414, Aug 2008.

\bibitem{Landau:1980}
L.~D. Landau, E.~M. Lifshitz, and L.~P. Pitaevskiı̆,
  \emph{\BIBforeignlanguage{English}{Statistical physics / by L.D. Landau and
  E.M. Lifshitz ; translated from the Russian by J.B. Sykes and M.J.
  Kearsley}}, 3rd~ed.\hskip 1em plus 0.5em minus 0.4em\relax Oxford; New York :
  Pergamon Press, 1980, translation of Statisticheskai︠a︡ fizika.

\bibitem{Newman:1999}
M.~E.~J. Newman and G.~T. Barkema, \emph{\BIBforeignlanguage{English}{Monte
  Carlo methods in statistical physics / M.E.J. Newman and G.T.
  Barkema}}.\hskip 1em plus 0.5em minus 0.4em\relax Oxford : Clarendon Press,
  1999.

\bibitem{Onsager:1944}
L.~Onsager, ``Crystal statistics. i. a two-dimensional model with an
  order-disorder transition,'' \emph{Phys. Rev.}, vol.~65, pp. 117--149, Feb
  1944.

\bibitem{Edwards:1975}
S.~F. Edwards and P.~W. Anderson, ``\BIBforeignlanguage{English}{Theory of spin
  glasses},'' \emph{\BIBforeignlanguage{English}{J. Phys. F}}, 1975.

\end{thebibliography}
\end{document}